\documentclass[9pt]{IEEEtran}
\IEEEoverridecommandlockouts
\usepackage{cite}
\usepackage{amsmath,amssymb,amsfonts}
\usepackage{algorithmic}
\usepackage{graphicx}
\usepackage{adjustbox}
\usepackage{textcomp}
\usepackage{hyperref}
\usepackage{booktabs}
\usepackage{multirow}
\usepackage{xurl}
\usepackage{makecell}
\usepackage{rotating}
\usepackage{float}
\usepackage{siunitx} 
\usepackage{xcolor}
\usepackage{cleveref}
\usepackage{adjustbox}
\def\BibTeX{{\rm B\kern-.05em{\sc i\kern-.025em b}\kern-.08em
    T\kern-.1667em\lower.7ex\hbox{E}\kern-.125emX}}
\usepackage[most]{tcolorbox}      
\tcbuselibrary{listings,skins,breakable}
\usepackage{listings}             
\usepackage{xcolor}               

\lstdefinestyle{pennycode}{
    basicstyle      = \ttfamily\scriptsize,  
    keywordstyle    = \color{blue},
    commentstyle    = \color{green!60!black},
    stringstyle     = \color{red},
    numbers         = left,
    numberstyle     = \tiny\color{gray},
    stepnumber      = 1,
    numbersep       = 5pt,
    backgroundcolor = \color{white},
    frame           = single,
    rulecolor       = \color{black!30},
    tabsize         = 4,
    captionpos      = b,
    breaklines      = true,
    breakatwhitespace = true,
    showstringspaces  = false
}
\usepackage[compact]{titlesec}

\titlespacing\section{0pt}{0.3\baselineskip}{0.2\baselineskip}
\titlespacing\subsection{0pt}{0.2\baselineskip}{0.1\baselineskip}
\titlespacing\subsubsection{0pt}{0.1\baselineskip}{0.1\baselineskip}

\usepackage{inconsolata}   
\usepackage{textcomp}      
\usepackage{pifont}        

\newcommand{\cmark}{\textcolor{green!60!black}{\ding{51}}} 
\newcommand{\xmark}{\textcolor{red}{\ding{55}}} 

\begin{document}

\title{PennyCoder: Efficient Domain-Specific LLMs for PennyLane-Based Quantum Code Generation}

\author{
\IEEEauthorblockN{Abdul Basit\textsuperscript{1}, Minghao Shao\textsuperscript{1}, Muhammad Haider Asif\textsuperscript{1,2}, Nouhaila Innan\textsuperscript{1,2}, Muhammad Kashif\textsuperscript{1,2}, Alberto Marchisio\textsuperscript{1,2},\\ Muhammad Shafique\textsuperscript{1,2}\\}
\IEEEauthorblockA{\textit{\textsuperscript{1}eBRAIN Lab, Division of Engineering} \textit{New York University (NYU) Abu Dhabi}, Abu Dhabi, UAE\\}
\textit{\textsuperscript{2}Center for Quantum and Topological Systems (CQTS), NYUAD Research Institute, New York University Abu Dhabi}, UAE\\
\{abdul.basit, shao.minghao, ma8183, nouhaila.innan, muhammadkashif, alberto.marchisio, muhammad.shafique\}@nyu.edu 
\vspace{-10pt}
}


\maketitle

\begin{abstract}

The growing demand for robust quantum programming frameworks has unveiled a critical limitation: current large language model (LLM) based quantum code assistants heavily rely on remote APIs, introducing challenges related to privacy, latency, and excessive usage costs. Addressing this gap, we propose PennyCoder, a novel lightweight framework for quantum code generation, explicitly designed for local and embedded deployment to enable on-device quantum programming assistance without external API dependence. PennyCoder leverages a fine-tuned version of the LLaMA 3.1-8B model, adapted through parameter-efficient Low-Rank Adaptation (LoRA) techniques combined with domain-specific instruction tuning optimized for the specialized syntax and computational logic of quantum programming in PennyLane, including tasks in quantum machine learning and quantum reinforcement learning. Unlike prior work focused on cloud-based quantum code generation, our approach emphasizes device-native operability while maintaining high model efficacy. We rigorously evaluated PennyCoder over a comprehensive quantum programming dataset, achieving 44.3\% accuracy with our fine-tuned model (compared to 33.7\% for the base LLaMA 3.1-8B and 40.1\% for the RAG-augmented baseline), demonstrating a significant improvement in functional correctness.

\end{abstract}

\begin{IEEEkeywords}
Quantum Programming, Large Language Models, Edge AI, PennyLane Code Generation, Domain-Specific Fine-Tuning, Retrieval-Augmented Generation (RAG).
\end{IEEEkeywords}

\section{Introduction}
Quantum computing is rapidly evolving from a theoretical pursuit to a practical technology, propelled by advances in both hardware and software. Milestones such as Google’s 105-qubit Willow processor~\cite{googleWillow2024}, IBM’s 1,000-qubit Condor~\cite{ibmCondor2023}, and Microsoft’s Majorana 1 device~\cite{microsoftMajorana2024} showcase growing capabilities in quantum hardware, driving demand for equally robust software tools. Frameworks like PennyLane~\cite{bergholm2018pennylane} have facilitated quantum algorithm design and execution, particularly in emerging applications such as Quantum Machine Learning (QML)~\cite{zaman2023survey} and Quantum Reinforcement Learning (QRL)~\cite{alomari2025survey}. However, there remains a significant gap in intelligent programming support for such specialized quantum paradigms.

Simultaneously, Large Language Models (LLMs) have transformed code generation across domains, including software engineering~\cite{chen2021evaluating}, hardware design~\cite{chakraborty2023hardwaregpt}, robotics~\cite{shinn2023reflexion}, and art~\cite{ramesh2022hierarchicaltextconditionalimagegeneration}. These successes are often enabled by domain-specific tuning, allowing LLMs to manage specialized syntax and logic with improved accuracy.

Despite these advances, quantum code generation remains underexplored. Tools like IBM’s Qiskit Assistant~\cite{qiskitAssistant} and RAG-based solutions~\cite{Lewis2020RetrievalAugmented, Kharkov2024BraketRAG, PennyLang} have demonstrated early promise, yet they largely depend on remote APIs (e.g., Codex, GPT-4), introducing latency, cost, privacy, and deployment constraints. Moreover, the intricate syntax and decision-making structures of quantum circuits designed for QRL, such as agent-based feedback, reward propagation, and dynamic Hamiltonian encoding, often exceed the capabilities of general-purpose models, even when augmented with retrieval.

Current approaches fall into two broad categories: (i) \textit{General Foundation Models}, which offer moderate performance but suffer from high hallucination rates; and (ii) \textit{RAG-Augmented Models}, which enhance contextual grounding at the cost of external dependencies. Neither approach sufficiently addresses the need for lightweight, locally deployable quantum programming assistants that can handle the evolving complexity of QRL environments.

Our empirical evaluation highlights this domain gap: baseline models like LLaMA 3.1-8B achieve only \textbf{33.7\%} accuracy, while RAG-augmented variants reach \textbf{40.1\%} on PennyLang \cite{PennyLang} dataset tasks. These limitations underscore the need for tailored solutions to meet the functional correctness requirements of quantum programming (see Section~\ref{sec:experiments}).

\begin{figure}[t]
    \centering
    \includegraphics[width=0.50\textwidth]{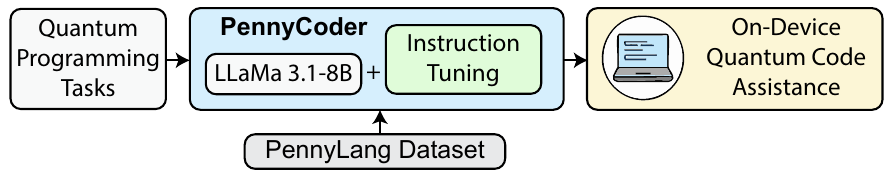}
    \vspace{-0.4cm}
    \caption{PennyCoder system, a lightweight on-device quantum code generation assistant. User queries are processed locally by a LoRA-tuned LLaMA 3.1-8B model, enhanced through domain-specific instruction tuning on the PennyLang dataset. The system outputs valid PennyLane-compatible quantum code and eliminates the need for external APIs by supporting embedded, privacy-conscious deployment.}
    \label{fig:pennycoder_introduction}
\end{figure}

To address this challenge, we introduce \emph{PennyCoder}, a lightweight, domain-adapted LLM framework designed for quantum code generation. PennyCoder integrates two key components: (1) a domain-specific model fine-tuned on the PennyLang dataset using instruction tuning; and (2) an efficient deployment strategy using Low-Rank Adaptation (LoRA). We further investigate decoding strategies (e.g., temperature, nucleus sampling) to improve output fidelity. Our contributions are as follows:
\begin{itemize}
\item[1.] We present \textit{PennyCoder}, a novel framework for quantum code generation that combines instruction fine-tuning with LoRA-based efficient adaptation.
\item[2.] We conduct a comprehensive evaluation demonstrating significant improvements over baseline and RAG-augmented models.
\item[3.] We analyze the impact of decoding hyperparameters on quantum code generation, offering insights for future system design.
\item [4.] We showcase the effectiveness of \textit{PennyCoder} on different use cases, including quantum algorithms, QML, and QRL tasks.
\end{itemize}

PennyCoder achieves an accuracy of \textbf{44.32\%} on tasks from the PennyLang dataset using LLaMA 3.1-8B as the foundation model, substantially outperforming both baseline and RAG-augmented models.

\section{Background and Related Work}
\label{sec:background}

\subsection{Large Language Models (LLMs) for Code Generation}
Large Language Models (LLMs) have shown strong performance in general-purpose code generation, with models like Codex~\cite{chen2021evaluating}, StarCoder~\cite{li2023starcoder}, and IBM's Granite~\cite{qiskit_ml} trained on diverse programming corpora. Benchmarks such as HumanEval~\cite{chen2021evaluating} and MBPP~\cite{austin2021program} have facilitated evaluation. However, their generalizability to niche quantum domains remains limited~\cite{watson2022qhack}, particularly when addressing quantum learning tasks such as Quantum Reinforcement Learning.



Prior research has explored the application of LLMs to quantum programming, primarily focusing on IBM's Qiskit framework. The Qiskit Code Assistant was fine-tuned to improve accuracy in quantum code generation tasks~\cite{GarciaAlmeida2024QiskitBlog}. Additionally, Vishwakarma et al. introduced Qiskit HumanEval, a benchmark designed to assess LLM-generated Qiskit code~\cite{Vishwakarma2024QiskitHumanEval}. However, despite PennyLane’s increasing adoption in QML, there has been limited research on evaluating LLMs for PennyLane-based quantum programming~\cite{basit2025qhackbench}. Our work fills this gap by systematically benchmarking LLM performance on PennyLane tasks using real-world coding challenges.

\subsection{Quantum Reinforcement Learning and PennyLane}
QRL is an emerging area where reinforcement learning paradigms intersect with quantum computing, enabling agents to interact with quantum environments to learn optimal policies~\cite{chen2024introduction, jerbi2023powerlimitationslearningquantum, chen2022quantum}. It requires efficient modeling of quantum states, unitaries, and reward-based adaptation across quantum circuits. PennyLane~\cite{bergholm2018pennylane}, developed by Xanadu, is particularly suited for this due to its hybrid execution model and support for automatic differentiation in variational quantum algorithms. These features make it a key tool for implementing QML and QRL pipelines, yet robust tools for intelligent code assistance in such contexts are still lacking.


\subsection{Domain Adaptation and RAG in Quantum Programming}

LLM adaptation strategies for quantum domains include both parameter-efficient fine-tuning and retrieval-augmented generation (RAG). While RAG has been used to extend LLM capabilities for dynamic environments like AWS Braket~\cite{Kharkov2024BraketRAG}, such methods introduce latency and external dependency issues, which limit practical deployment in privacy-sensitive QRL applications.

\begin{figure}[ht]
    \centering
                    \includegraphics[width=0.5\textwidth]{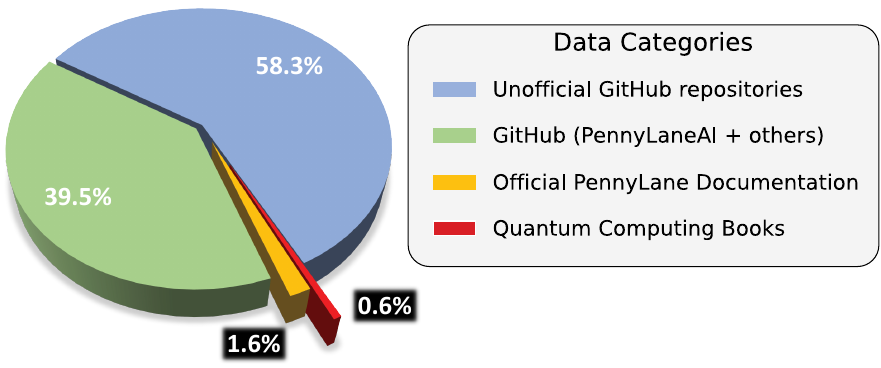}
    \caption{Composition of the \textit{PennyLang} dataset (3,347 samples).}
    \label{fig:pennylang_composition}
\end{figure}

\textit{PennyLang}~\cite{PennyLang}, a curated instruction-code dataset for PennyLane, provides the necessary grounding for adapting LLMs to quantum programming tasks, particularly for QML and QRL. It contains natural language prompts and corresponding PennyLane code samples, drawn from documentation, community repositories, and quantum textbooks. This structure enables both direct fine-tuning and retrieval-based augmentation, enhancing LLM accuracy in quantum-specific domains. Its composition is shown in Figure~\ref{fig:pennylang_composition}.

 \begin{figure*}[ht]
    \centering
    \includegraphics[width=0.95\textwidth]{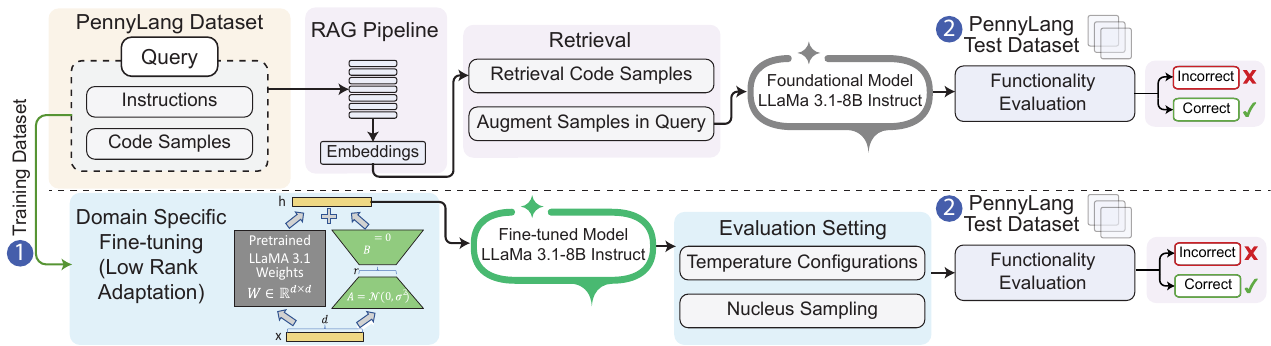}
    \caption{Overview of the PennyCoder framework. The PennyLang dataset, structured in an instruction-code format, is used both for direct domain-specific fine-tuning of a foundation model (using Low-Rank Adaptation) and for building a retrieval corpus in the RAG pipeline. During training, fine-tuning adapts a LLaMA 3.1-8B model to PennyLane quantum programming tasks. During evaluation, the system supports both direct generation and RAG to enhance few-shot generalization. Generated outputs are evaluated for functional correctness against the PennyLang test set, using controlled decoding settings such as temperature scaling and nucleus (top-$p$) sampling.}
    \label{fig:pennycoder_methodology}
\end{figure*}

\subsection{Comparison of Related Work}
Table~\ref{tab:related_work_comparison} summarizes representative prior works and compares them with PennyCoder across key axes including backend framework, support for QRL/QML, deployment modality, and local inference capability.

\begin{table}[ht]
\centering
\scriptsize
\caption{Comparison of Related Work on LLM-based Quantum Code Gen}
\label{tab:related_work_comparison}
\begin{tabular}{p{2.2cm}p{0.6cm}p{1cm}p{0.6cm}p{0.6cm}p{1.3cm}}
\toprule
\textbf{Method} & \textbf{Year} & \textbf{Backend} & \textbf{QML} & \textbf{QRL} &  \textbf{Fine-Tuned} \\
\midrule
Codex + Qiskit \cite{chen2021evaluating} & 2021 & Qiskit & \cmark & \xmark &  \xmark \\
Qiskit Assistant~\cite{qiskitAssistant} & 2023 & Qiskit & \cmark & \cmark &  \cmark \\
Braket RAG~\cite{Kharkov2024BraketRAG} & 2024 & Braket & \cmark & \xmark & \xmark \\
PennyLang~\cite{PennyLang} & 2024 & PennyLane & \cmark & \cmark &  \xmark \\
\textbf{PennyCoder (Ours)} & 2025 & PennyLane & \cmark & \cmark &  \cmark \\
\bottomrule
\end{tabular}
\end{table}

\section{PennyCoder Framework}\label{sec:methodology}

The \emph{PennyCoder} framework is designed as a lightweight, domain-adapted solution for quantum code generation within the PennyLane ecosystem, with a particular focus on enabling Quantum Reinforcement Learning (QRL) pipelines. Unlike prior cloud-dependent approaches, PennyCoder is optimized for \textit{local and embedded deployment}, facilitating on-device inference without relying on external APIs (e.g., OpenAI, DeepSeek).


The system is composed of three modules: (1) a domain-specific instruction-response dataset curated from PennyLane programming tasks, (2) a parameter-efficient fine-tuning pipeline based on Low-Rank Adaptation (LoRA) techniques, and (3) an additional Retrieval-Augmented Generation (RAG) module to enhance model robustness for long-tail tasks. Figure~\ref{fig:pennycoder_methodology} illustrates the PennyCoder framework.


\subsection{Instruction Tuning with Domain-Specific Data}
\label{sec:instruction_tuning}


We leverage an instruction-response formatted corpus derived from curated PennyLang dataset \cite{PennyLang}. Each data point consists of a natural language instruction describing a quantum task, paired with its corresponding PennyLane code implementation. This structure enables effective supervised fine-tuning, improving the model’s ability to map textual quantum programming queries into executable code. The instruction-response format enables fine-tuning using simple supervised loss objectives such as cross-entropy minimization, allowing efficient and stable optimization even with limited computational resources.

\subsubsection{Fine-tuning Configuration}

The fine-tuning was performed using the PennyLang dataset under a parameter-efficient training setup. A learning rate of $1\times10^{-6}$ was used for 2 epochs on a single NVIDIA A100 80GB GPU. Due to limited batch size (1), gradient accumulation with 4 steps was employed. Mixed-precision training (\texttt{fp16}) was enabled to further optimize memory usage and computational efficiency. A maximum input token length of 15,000 was set to accommodate long instruction-code pairs common in QML routines.




\subsubsection{Low-Rank Adaptation (LoRA)}
To support parameter-efficient fine-tuning, we apply LoRA to key transformer layers. Given an attention weight matrix $W \in \mathbb{R}^{d \times d}$:
\[
W' = W + \Delta W, \quad \Delta W = A B,
\]
where $A \in \mathbb{R}^{d \times r}$ and $B \in \mathbb{R}^{r \times d}$, with $r \ll d$.


In PennyCoder:
\begin{itemize}
    \item Rank $r = 8$ was empirically selected.
    \item LoRA was applied to query and value matrices of self-attention layers.
    \item A dropout rate of 0.05 was used.
    \item The AdamW optimizer was used with the same learning rate.
\end{itemize}

This strategy enables adaptation to QML/QRL tasks without full model retraining, reducing memory cost and enhancing portability to QRL simulation environments and hardware-constrained quantum SDKs.


\subsection{Retrieval-Augmented Generation (RAG) for Code Generalization}
\label{sec:rag_module}

While instruction fine-tuning provides a strong inductive bias, long-tail QRL scenarios (e.g., dynamic circuit mutation, custom reward functions) benefit from contextual retrieval. We implement a RAG module based on~\cite{Lewis2020RetrievalAugmented}, tuned for PennyLang.


The retrieval process proceeds as follows:
\begin{enumerate}
    \item Embed the user query into a dense vector space using an Instructor-Large encoder.
    \item Retrieve top-$k$ most similar instructions from the PennyLang corpus using cosine similarity in the vector space.
    \item Concatenate the retrieved instruction-code pairs into the model input context window.
\end{enumerate}


This mechanism improves output consistency in QML/QRL applications where direct instruction supervision is sparse. It complements the foundation model by injecting retrieval content.

PennyCoder combines the strengths of instruction fine-tuning and lightweight parameter-efficient adaptation techniques to deliver a robust, locally deployable LLM solution for PennyLane quantum programming. By addressing both data scarcity and compute constraints, PennyCoder provides a practical path forward for scalable, privacy-preserving quantum code generation.

\section{Experiment Settings}
\label{sec:experiments}

\subsection{Dataset}
We use the PennyLang dataset~\cite{PennyLang} for both instruction fine-tuning and retrieval-augmented generation (RAG). The dataset contains 3,347 curated samples sourced from GitHub repositories, quantum computing textbooks, and the official PennyLane documentation. Each sample was manually verified for accuracy and relevance to quantum programming.

To adapt our model to the quantum domain, we apply instruction fine-tuning (see Section~\ref{sec:methodology}) using LoRA for parameter-efficient training. Each example is augmented with high-quality instructions generated by LLMs to contextualize the task and guide model behavior. This process enhances output relevance and accuracy in quantum-specific coding scenarios.

\subsection{Benchmark and Metrics}
To evaluate PennyCoder, we created a custom benchmark comprising of 264 tasks for PennyLane code generation inspired by the Qiskit HumanEval benchmark~\cite{Vishwakarma2024QiskitHumanEval}, covering a range of difficulty levels and quantum topics such as circuit construction, algorithm implementation, variational quantum eigensolvers, and quantum machine learning. Tasks are drawn from prior competitions and categorized into foundational and advanced application domains.

\begin{figure}[ht]
    \centering
    \includegraphics[width=\linewidth]{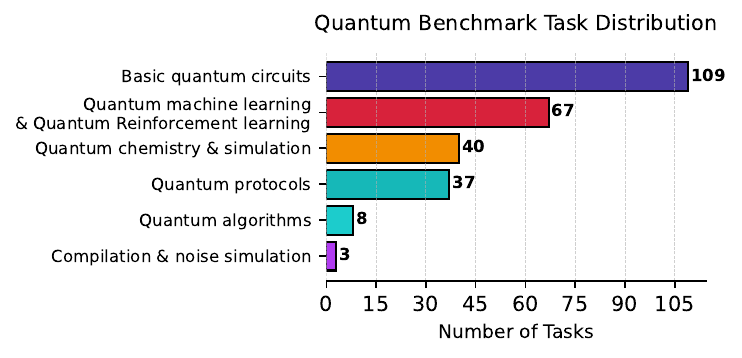}
    \vspace{-0.6cm}
    \caption{Distribution of benchmark tasks across quantum programming categories. Basic circuits represent the largest share (41\%). Each task is assigned to its most relevant category.}
    \label{fig:rag_results}
\end{figure}

To measure performance, we use \textit{Pass@k} metrics, evaluating syntactic validity and functional correctness across $k = 1, 3, 5$ model completions. This accounts for output variability and assesses the probability of generating a correct solution within $k$ attempts. All outputs are manually verified to ensure reliable evaluation.

\subsection{Implementation Details}
Our experiments use LLaMA 3.1 8B~\cite{llama31} as the base model. To explore decoding dynamics, we conduct a grid search over temperature and nucleus sampling (top-$p$) with values $\{0, 0.5, 1.0\}$. This analysis reveals the impact of generation hyperparameters on output determinism and creativity, informing optimal deployment configurations.

\section{Results}

\subsection{PennyCoder Evaluation}


\Cref{tab:pennycoder_comparison} presents a comparative evaluation of PennyCoder, which employs our proposed PennyCoder, against two baseline approaches: the base LLaMA model and LLaMA with RAG.

PennyCoder demonstrates substantial improvements over the base LLaMA model. With 117 successes versus LLaMA's 89 (a 31.5\% increase), PennyCoder significantly reduces failures from 175 to 147 (a 16\% reduction). This performance boost translates to an accuracy improvement from 33.71\% to 44.32\%, representing an absolute gain of 10.61 percentage points.
When compared to the RAG-enhanced baseline, PennyCoder maintains its superiority. While LLaMA + RAG achieves 106 successes and 40.15\% accuracy, PennyCoder surpasses it with 11 additional successes and a 4.17 percentage point accuracy increase.

These results establish PennyCoder as an effective approach among the three configurations. By achieving consistent performance gains across all metrics without requiring external retrieval mechanisms, our method demonstrates that low-rank adaptation fine-tuning successfully enhances model capabilities in the quantum computing domain.

\begin{table}[htbp]
\centering
\caption{Performance of PennyCoder (with T=0.5 and top\_p=0.5) compared with our baselines.}
\begin{tabular}{cccc}
\toprule
& \textbf{{PennyCoder (Ours)}} & \textbf{LLaMA} & \textbf{LLaMA + RAG}\\
\midrule
\textbf{Success} & 117 & 89 & 106 \\
\textbf{Failed} & 147 & 175 & 158 \\
\textbf{Accuracy} (\%) & \textbf{44.32} & 33.71 & 40.15 \\
\bottomrule
\end{tabular}
\label{tab:pennycoder_comparison}
\end{table}

\subsection{Category-wise Performance}

\begin{figure}[t]
    \centering
    \includegraphics[width=1\linewidth]{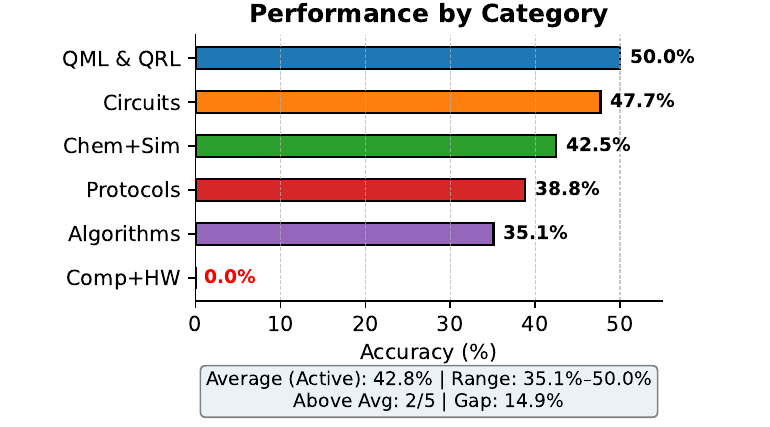}
    \caption{Category-wised accuracy analysis. Explanation of task category abbreviation: Circuits=Basic quantum circuits; Comp+HW=Compilation, noise simulation \& hardware mapping; Algorithms=Quantum algorithms, QML\&QRL; Chem+Sim=Quantum chemistry \& simulation; Protocols=Quantum protocols.}
    \label{fig:acc_radar}
\end{figure}

\Cref{fig:acc_radar} reveals variable performance across quantum computing categories. In Basic Quantum Circuits, the largest test set with 109 cases, the model achieved moderate proficiency with 52 successes against 57 failures. Similarly, in Quantum Machine Learning (67 cases), the model struggled with complex applications, producing only 26 successful outcomes versus 41 failures.

Performance declined further in specialized domains. Quantum Protocols showed limited success with 13 passes out of 37 tests, while Quantum Chemistry and Simulation yielded 17 successes from 40 tests. The Quantum Algorithms category, though smallest at 8 cases, resulted in an even split of 4 successes and 4 failures. Most concerning, PennyCoder failed all 3 Compilation and Noise Simulation tests, indicating no capability in hardware-level or noise-aware tasks.

Accuracy metrics show in \Cref{fig:acc_radar} reinforce these patterns. While the model achieved 50\% accuracy in algorithms (limited sample size), it performed relatively well in basic circuits at 47.71\%. Performance decreased progressively in chemistry and simulation (42.5\%), machine learning (38.81\%), and protocols (35.14\%), with compilation and noise simulation at 0\%.

These results suggest PennyCoder's strengths lie in basic circuit construction and algorithmic reasoning, while significant challenges emerge in complex and specialized domains. The inaccuracies in the noise simulation suggest that dedicated adjustments are required to improve the performance for this category. Moving forward, both enhanced training protocols and architectural improvements will be essential to address these domain-specific quantum computing challenges.


\subsection{Impact of Temperature and Nucleus Sampling Values}


\Cref{fig:hp_results} summarizes the pass rate across different temperature ($T$) and nucleus sampling (top-$p$) configurations. For $T=0$, the pass rate remained consistent at approximately 42.42\% across top-$p$ values, with a slight improvement to 43.18\% when top-$p=1.0$. When setting $T=0.5$, the model achieved its highest accuracy of 44.32\% at top-$p=0.5$, suggesting that moderate randomness in generation improves model output. However, at $T=0.5$ and top-$p=1.0$, performance dropped to 37.88\%, indicating that excessive sampling diversity can negatively impact generation quality. Notably, setting $T=1$ resulted in a consistent decline in accuracy—38.64\%, 33.33\%, and 5.68\% for top-$p=0.0$, $0.5$, and $1.0$ respectively—reinforcing the notion that high temperature without proper sampling control leads to unstable and degraded model behavior. These results underscore the importance of fine-tuning decoding parameters, with $T=0.5$ and top-$p=0.5$ standing out as the most effective combination for maximizing performance.

\begin{table}[ht!]
\centering
\caption{Pass rate of different hyperparameter setup \{T=[0, 0.5, 1]; top\_p=[0, 0.5, 1]\}}
\label{fig:hp_results}
\footnotesize
\setlength{\tabcolsep}{2.5pt}
\begin{adjustbox}{max width=\linewidth}
\begin{tabular}{l*{9}{S[table-format=3.0,tight-spacing=true]}}
\toprule
\multicolumn{1}{c}{} & 
\multicolumn{3}{c}{\textbf{T=0}} & 
\multicolumn{3}{c}{\textbf{T=0.5}} & 
\multicolumn{3}{c}{\textbf{T=1}} \\
\cmidrule(lr){2-4} \cmidrule(lr){5-7} \cmidrule(lr){8-10}
\multicolumn{1}{c}{\textbf{top\_p}} & 
\textbf{0.0} & \textbf{0.5} & \textbf{1.0} & 
\textbf{0.0} & \textbf{0.5} & \textbf{1.0} & 
\textbf{0.0} & \textbf{0.5} & \textbf{1.0} \\
\midrule
\textbf{Success} & 112 & 112 & 114 & 112 & \textbf{117} & 100 & 102 & 88 & 15 \\
\textbf{Failed}  & 152 & 152 & 150 & 152 & \textbf{147} & 164 & 162 & 176 & 249 \\
\multicolumn{1}{l}{\textbf{Accuracy (\%)}} & 
\multicolumn{1}{c}{42.42} & \multicolumn{1}{c}{42.42} & \multicolumn{1}{c}{43.18} & 
\multicolumn{1}{c}{42.42} & \multicolumn{1}{c}{\textbf{44.32}} & \multicolumn{1}{c}{37.88} & 
\multicolumn{1}{c}{38.64} & \multicolumn{1}{c}{33.33} & \multicolumn{1}{c}{5.68} \\
\bottomrule
\end{tabular}
\end{adjustbox}
\end{table}





\section{Discussion}

\subsection{Case Study I: PennyCoder vs. Foundation Model vs. RAG}
The task is to construct a QRL agent using a two-qubit parameterized circuit to choose between left or right in a 1D environment.

\begin{tcolorbox}[colback=green!3!white,colframe=green!90!black,
                  title=PennyCoder Output,
                  breakable]
\scriptsize
\textbf{Limitations Identified:}
\begin{itemize}
    \item No state encoding in the quantum circuit.
    \item No environment interaction loop.
    \item Manual parameter updates without gradients.
    \item Limited learning mechanism.
\end{itemize}

\begin{lstlisting}[caption=PennyCoder Implementation]
import pennylane as qml
from pennylane import numpy as np
import torch
class QRLAgent:
    def __init__(self):
        self.device = qml.device('default.qubit', wires=2)
        self.circuit = qml.QNode(self._circuit, self.device)
        self.params = qml.numpy.array([0.1, 0.1])
    def _circuit(self, weights):
        qml.RX(weights[0], wires=0)
        qml.RX(weights[1], wires=1)
        qml.CNOT(wires=[0, 1])
        qml.RY(weights[0], wires=0)
        qml.RY(weights[1], wires=1)
        return qml.expval(qml.PauliZ(0))
    def act(self, state):
        prob = self.circuit(self.params)
        if prob > 0.5:
            return 1
        else:
            return 0
    def update(self, reward, next_state):
        # update the parameters using a simple gradient descent
        loss = (reward + 1) * (self.act(next_state) - 0.5)
        self.params -= 0.1 * loss
# example usage
agent = QRLAgent()
state = 0
reward = 1
next_state = 1
agent.update(reward, next_state)
print(agent.act(state))  # prints 0 or 1 depending on the updated parameters

\end{lstlisting}

\textbf{Description:}
The agent meets the basic requirements but lacks dynamic state encoding, proper reinforcement learning feedback, and gradient-based optimization, which limits its effectiveness in real QRL scenarios. 
\end{tcolorbox}

\begin{tcolorbox}[colback=red!3!white,colframe=red!90!black,
                  title=LLaMA 3.1‑8B Foundational Model (Wrong Output),
                  breakable]
\scriptsize
\textbf{Limitations Identified:}
\begin{itemize}
    \item No training mechanism or weight optimization.
    \item State is not encoded into the quantum circuit.
    \item Actions are selected randomly.
    \item Circuit output is not used to guide learning.
\end{itemize}

\begin{lstlisting}[caption=LLaMA 3.1‑8B Implementation (Baseline)]
import gym
@qml.qnode(qml.device("default.qubit", wires=2))
def circuit(params):
    qml.RX(params[0], wires=0)
    qml.RY(params[1], wires=0)
    qml.RZ(params[2], wires=0)
    qml.RX(params[3], wires=1)
    qml.RY(params[4], wires=1)
    qml.RZ(params[5], wires=1)
    qml.CNOT(wires=[0, 1])
    return qml.expval(qml.PauliZ(wires=1))
class QRLAgent:
    def __init__(self, env, device):
        self.env = env
        self.device = device
        self.params = [qml.numpy.array([0.1]*6)]
        self.qnode = qml.qnode(device, circuit=circuit)
    def act(self):
        self.qnode(self.params)
        prob_left = qml.math.probability(self.qnode, 0)
        prob_right = qml.math.probability(self.qnode, 1)
        if prob_left > prob_right:
            return 0
        else:
            return 1
    def update(self, reward, new_state):
        self.params = qml.math.adam(self.params, reward, lr=0.01)
env = gym.make('CartPole-v1')
device = qml.device("default.qubit", wires=2)
agent = QRLAgent(env, device)
for episode in range(10):
    state = env.reset()
    done = False
    while not done:
        action = agent.act()
        new_state, reward, done, _ = env.step(action)
        agent.update(reward, new_state)
        env.render()  # Render the environment
    print(f'Episode: {episode}, Reward: {reward}')  # Print the reward for each episode
\end{lstlisting}

\textbf{Description:}
Attempts to do the task, but the implementation lacks essential components such as learning dynamics, state encoding, and parameter updates, limiting its effectiveness in reinforcement learning tasks.
\end{tcolorbox}

\begin{tcolorbox}[colback=red!3!white,colframe=red!90!black,
                  title=LLaMA 3.1‑8B + RAG (Wrong Output),
                  breakable]
\scriptsize
\textbf{Limitations Identified:}
\begin{itemize}
        \item No training or weight update mechanism.
    \item The circuit output overwrites the parameter tensor.
    \item State is not encoded into the quantum circuit.
    \item Random action selection without policy guidance.
\end{itemize}

\begin{lstlisting}[caption=LLaMA 3.1‑8B + RAG Implementation (Partial)]
import pennylane as qml
from pennylane import numpy as np
import torch
# Define the QRL agent
class QRLAgent:
    def __init__(self, device, num_qubits, num_layers):
        self.device = device
        self.num_qubits = num_qubits
        self.num_layers = num_layers
        self.circuit = self._create_circuit()
    def _create_circuit(self):
        # Define the QRL circuit
        @qml.qnode(self.device)
        def circuit(weights, action):
            # Apply a parameterized circuit to choose between left or right
            for i in range(self.num_layers):
                qml.RX(weights[2 * i], wires=0)
                qml.RX(weights[2 * i + 1], wires=1)
                qml.CNOT(wires=[0, 1])
            if action == 0:  # Left
                qml.RX(np.pi, wires=0)
            else:  # Right
                qml.RX(-np.pi, wires=0)
            return qml.expval(qml.PauliZ(0))
        return circuit
    def choose_action(self, weights, action):
        # Choose between left or right using the QRL circuit
        return self.circuit(weights, action)
# Create a QRL agent
device = qml.device("default.qubit", wires=2)
num_qubits = 2
num_layers = 2
agent = QRLAgent(device, num_qubits, num_layers)
# Define the environment
class Environment:
    def __init__(self):
        self.state = 0
    def step(self, action):
        # Update the environment state based on the action
        if action == 0:  # Left
            self.state -= 1
        else:  # Right
            self.state += 1
        return self.state
# Create an environment
env = Environment()
# Train the QRL agent
weights = torch.randn(4)
for i in range(100):
    action = 0 if np.random.rand() < 0.5 else 1
    state = env.step(action)
    reward = 1 if state == 0 else -1
    weights = agent.choose_action(weights, action)
    print(f"Iteration {i+1}, Action: {action}, Reward: {reward}")
\end{lstlisting}

\textbf{Description:}
Attempts to do the task, 
but, the lack of state encoding, absence of a learning mechanism, and incorrect handling of circuit output limit the agent's ability to learn or adapt over time. 
\end{tcolorbox}

\subsection{Case Study II: PennyCoder vs. Foundation Model vs. RAG}
The task is to create a quantum circuit that prepares a state for superdense coding, and provide it as a QNode.
\begin{tcolorbox}[colback=green!3!white,colframe=green!90!black,
                  title=PennyCoder Output (Working),
                  breakable]
\scriptsize
\textbf{Limitations Identified:}
\begin{itemize}
    \item Uses the default simulator; no hardware optimisation
    \item Returns the full state vector (may be large for bigger systems)
\end{itemize}

\begin{lstlisting}[caption=PennyCoder Implementation]
import pennylane as qml

def superdense_coding_circuit():
    """Design a circuit that prepares a state suitable for superdense coding. 
    Use a Bell state preparation and additional gates if necessary. Return the QNode."""

    @qml.qnode(qml.device('default.qubit', wires=2))
    def circuit():
        qml.Hadamard(wires=0)
        qml.CNOT(wires=[0, 1])
        return qml.state()

    return circuit

# Example usage
circuit = superdense_coding_circuit()
print(circuit())
\end{lstlisting}

\textbf{Description:}
Creates a minimal two‑qubit Bell‑state circuit suitable for super‑dense coding
protocols. The function returns a ready‑to‑use QNode that outputs the full
quantum state.
\end{tcolorbox}

\begin{tcolorbox}[colback=red!3!white,colframe=red!90!black,
                  title=LLaMA 3.1‑8B Foundational Model (Compilation Errors),
                  breakable]
\scriptsize
\textbf{Limitations Identified:}
\begin{itemize}
    \item Introduces unused classical arguments \texttt{x, y}
    \item Calls \texttt{qml.PauliX(...).control(1)} – unsupported syntax
    \item Multiple redundant single‑qubit \texttt{PauliX} gates
    \item Returns \texttt{qml.state()} but gradient parameters are unused
\end{itemize}

\begin{lstlisting}[caption=LLaMA 3.1‑8B Implementation (Baseline)]
import pennylane as qml

def superdense_coding_circuit():
    """Design a circuit that prepares a state suitable for superdense coding. Use a Bell state preparation and additional gates if necessary. Return the QNode."""

    # Define the QNode
    @qml.qnode(qml.device("default.qubit", wires=2))
    def circuit(x, y):
        # Apply Hadamard gate to the first qubit
        qml.Hadamard(wires=0)
        # Apply Pauli-X gate to the first qubit
        qml.PauliX(wires=0)
        # Apply CNOT gate to the first and second qubits
        qml.CNOT(wires=[0, 1])
        # Apply Pauli-X gate to the second qubit
        qml.PauliX(wires=1)
        # Apply Pauli-X gate to the second qubit conditioned on the first qubit
        qml.PauliX(wires=1).control(1)(wires=[0, 1])
        return qml.state()

    return circuit

# Example usage
circuit = superdense_coding_circuit()
print(circuit(0, 0))
\end{lstlisting}

\textbf{Description:}
Attempts to extend the Bell‑state circuit but introduces unsupported controlled
operations and unused parameters, leading to runtime errors when executed.
\end{tcolorbox}

\begin{tcolorbox}[colback=red!3!white,colframe=red!90!black,
                  title=LLaMA 3.1‑8B + RAG (Compilation Errors),
                  breakable]
\scriptsize
\textbf{Limitations Identified:}
\begin{itemize}
    \item Tries to prepare basis state \textbf{outside} a QNode context
    \item Returns a bare \texttt{qml.qnode} object without a wrapped function
    \item Example call \texttt{circuit()} fails because no circuit body exists
\end{itemize}

\begin{lstlisting}[caption=LLaMA 3.1‑8B + RAG Implementation (Partial)]
import pennylane as qml
from pennylane import numpy as np

def superdense_coding_circuit():
    """Design a circuit that prepares a state suitable for superdense coding. Use a Bell state preparation and additional gates if necessary. Return the QNode."""

    # Prepare the Bell state
    qml.BasisState([0, 1], wires=[0, 1])

    # Apply a Hadamard gate to the first qubit
    qml.Hadamard(wires=0)

    # Apply a Pauli-X gate to the second qubit
    qml.PauliX(wires=1)

    # Return the QNode
    return qml.qnode(qml.device("default.qubit", wires=2))

# Example usage
circuit = superdense_coding_circuit()
print(circuit())
\end{lstlisting}

\textbf{Description:}
Moves towards using PennyLane primitives but places gate operations outside any
QNode scope and returns an uninitialised qnode, causing errors when invoked.
\end{tcolorbox}

\subsection{Limitation and Future Work}
While PennyCoder demonstrates promising results in LLM-aided quantum code generation through domain-adaptive model fine-tuning, several key challenges remain within our current framework. First, our training dataset comprises only 3,000 samples, raising questions about potential improvements through data augmentation or enhanced instruction generation for specific scenarios. Could expanding the dataset or refining instruction templates lead to better performance?

Second, our experiments utilize LLaMA 3.1 8B as the foundation model. As state-of-the-art language models continue to evolve, it would be valuable to evaluate how newer architectures perform within the PennyCoder framework. Given that many modern inference models, including OpenAI o3 \cite{openaio3mini}, Claude 3.7 \cite{claude37}, and DeepSeek R1 \cite{deepseekr1}, incorporate Chain-of-Thought reasoning, future work should explore how to enhance both the output quality and reasoning capabilities of these models when integrated with PennyCoder.

Based on these limitations, future work on the PennyCoder framework should focus on two key directions. First, advanced data augmentation techniques to address the constraints of limited training data availability can be explored. Second, following work can enhance the model's output quality and evaluation capabilities by incorporating agentic techniques, such as conversational feedback mechanisms and tool usage integration. These approaches will enable more sophisticated automated evaluation while improving the model's practical utility in quantum code generation.

\section{Conclusion}

In this work, we presented \textit{PennyCoder}, a lightweight and efficient framework for PennyLane-based quantum code generation. By fine-tuning a foundation model using Low-Rank Adaptation (LoRA) techniques and leveraging domain-specific instruction datasets, PennyCoder achieves significant improvements in functional correctness compared to both standard foundation models and retrieval-augmented methods. Our model maintains competitive accuracy while being deployable locally, thus addressing critical challenges of latency, privacy, and deployment feasibility often associated with remote API-based solutions.

Our experiments highlight that fine-tuning alone, when carefully applied, can outperform RAG-augmented approaches without the need for complex retrieval pipelines. Additionally, we demonstrate that careful calibration of generation hyperparameters (temperature and nucleus sampling) can further boost the success rate of quantum code generation. 

Future work will focus on expanding PennyCoder's capabilities for advanced quantum hardware-aware tasks, incorporating noise models, and enabling broader support for hybrid classical-quantum workflows. We believe PennyCoder represents an important step toward building scalable, private, and robust LLM-based quantum programming assistants.

\section*{Acknowledgment}
 This work was supported in part by the NYUAD Center for Quantum and Topological Systems (CQTS), funded by Tamkeen under the NYUAD Research Institute grant CG008, the NYUAD Center for CyberSecurity (CCS), funded by Tamkeen under the NYUAD Research Institute Award G1104, and the NYUAD High Performance Computing (HPC) center for providing necessary compute resources for the experiments.

\bibliographystyle{IEEEtran}
\bibliography{main}

\end{document}